\documentclass[12pt]{article}
\usepackage{graphics,graphicx}

\usepackage{txfonts}
\begin{document}

\title{Description of accretion
induced outflows from ultra-luminous sources to under-luminous AGNs}

\maketitle
\author{Shubhrangshu Ghosh$^{1,2,3}$, Banibrata 
Mukhopadhyay$^{2}$, Vinod Krishan$^{1,4}$,
Manoranjan Khan$^{3}$\\
1. Indian Institute of Astrophysics, Bangalore 560034, India; 
sghosh@iiap.res.in, vinod@iiap.res.in\\
2. Astronomy and Astrophysics Programme, Department of Physics,
Indian Institute of Science, Bangalore 560012, India; bm@physics.iisc.ernet.in\\ 
3. Department of Instrumentation Science \& Center for Plasma Studies, 
Jadavpur University, Kolkata 700032, India; mkhan@isc.jdvu.ac.in\\
4. Raman Research Institute, Bangalore 560080, India
}



\begin{abstract}

We study the energetics of the accretion-induced outflow and then plausible
jet around black holes/compact 
objects using a newly developed disc-outflow
coupled model. Inter-connecting dynamics of outflow and accretion 
essentially upholds the conservation laws. 
The energetics depend strongly on the viscosity parameter $\alpha$ and the cooling 
factor $f$ which exhibit several interesting features. The bolometric 
luminosities of ultra-luminous X-ray binaries (e.g. SS433) and family of highly luminous 
AGNs and quasars can be reproduced by the model under the super-Eddington accretion flows. 
Under appropriate conditions, low-luminous AGNs 
(e.g. Sagittarius $A^*$) also fit reasonably well with the luminosity corresponding
to a sub-Eddington accretion flow with $f\rightarrow 1$.  


\end{abstract}

keywords: accretion, accretion disc --- black hole physics --- 
X-rays: binaries --- galaxies: jets --- galaxies: nuclei 

\label{firstpage}

\section{Introduction}

Extremely high resolution observations of the powerful extragalactic
 double radio sources \cite{BR, BBR, FER, MIR} reveal that they are formed by
 well-collimated outflows or jets that continuously emerge from 
the nuclear region of the host active galaxies (AGNs) or 
quasars, believed to harbor supermassive black holes. Similarly 
micro-quasars \cite{MIRR94, MIRR98} 
discovered in recent times reveal that outflows are generated from stellar 
mass black holes (or black holes candidates).
The well-collimated outflow in SS433 observed for several decades, which is our galactic, persistent, 
super-critical accretor, is an well known evidence of
cosmic jet \cite{MAR84}. 
Further, highly collimated line jets are seen in young stellar objects 
\cite{MUN85}. Relativistic jets are also observed in neutron star 
low mass X-ray binaries (LMXBs) \cite{MF06}. 

The outflows/jets extract matter, energy and momentum from the accretion disc that forms around the 
compact object,  
and thus the dynamics of outflow leading to jet is intrinsically coupled with the accretion dynamics 
through the conservation laws. 
Also several observations (precisely the simultaneous observations of 
disc and jet; see e.g. Ghosh \& Mukhopadhyay 2009, and references therein) reveal that 
accretion processes 
and outflows are strongly correlated and they eventually control
the accretion process, precisely, the accretion dynamics in the vicinity 
of the central star. The relativistic 
outflowing matter, in the case of quasars or micro-quasars, should indeed come
 only from the inner region of the accretion disc. 
This is particularly suggestive as the quasars or the micro-quasars do not 
have an atmosphere of their own. However, most of the models of accretion disc and outflow/jet have been evolved separately, considering these two to be apparently dissimilar objects. The outflow/jet models have been
evolved over the years,  
from speculative ideas such as de Laval nozzles \cite{BR} to 
electrodynamic acceleration model \cite{BZ}, centrifugally 
driven outflows \cite{BP82, PN86}, etc. Moreover, MHD simulations of outflow/jet have been performed both in non-relativistic as well as relativistic 
limits, mostly in the Keplerian paradigm (e.g. Hawley \& Balbus 2002, Mizuno et al. 2006, Hawley \& Krolik 2006, and
references therein) to see how the matter gets deflected from the equatorial plane. Nevertheless, the definitive understanding of the origin of outflows/jets is sill unknown. 

The radiatively driven outflow or jet whose origin is better understood, 
can be envisaged when the accretion disc is highly radiation pressure 
dominated or precisely ``radiation trapped''. This is likely to occur when the accretion rate is super-Eddington 
or super-critical \cite{LOV94, BEG, FAB, GM09} 
as in ultra luminous X-ray (ULX) sources such as SS433 (with luminosity $\sim 10^{40}$ erg/s or 
so) \cite{FAB}; a prototype of ULXs in external galaxies having relativistic jets. The super-critical flows 
are optically thick and have a strong advective component \cite{LIP, OSH}.  
The unusual signatures of 
ULXs, compared to their counterparts, lie in their extreme high luminosity 
($L_{bol} \sim 10^{39-41}$ erg/s), strong spectral variability and their 
association with the actively star forming regions, that make their
nature a subject of controversy \cite{FAB, MUS}. Recently, an ultra-luminous 
accretion disc with a high kinetic luminosity 
radio jet has been discovered in quasar PKS~0743-67 \cite{PT},  
which indicates 
that the ultra-luminous sources may be extended to quasars or even 
AGNs. On the other extreme end, the under-luminous AGNs and quasars (e.g. Sagittarius $A^{*}$) 
had been described by the advection dominated accretion flow model (in short 
ADAF \cite{NY94}),  
where the flow is substantially sub-critical/sub-Eddington. The model hypothesized a 
plausible emergence of strong outflows/jets, which lead to reveal a strong 
inter-connection between the outflow and advection. 

Very few models exist which simultaneously deal with the accretion and outflow/jet 
dynamics on the same platform, both in analytical regime as well as through numerical simulations 
(see Ghosh \& Mukhopadhyay 2009, and references therein). The difficulty of simultaneous 
simulation of the disc and outflow arises due to the difference in time scales 
between accretion and outflow. 
Also the results strongly depend on the initial conditions \cite{UST}. 
The 2.5-dimensional disc-outflow coupled model given by Ghosh \& Mukhopadhyay (2009) has
been formulated by incorporating explicit information of the 
outflow in a fully analytical regime through a self-similar approach in a general advective paradigm, by 
upholding the conservation laws. It has been further shown that with mass, energy and momentum conservation, and a 
few scaling arguments, the 
correlation of the jet with the disc could be successfully modeled. The authors 
solved a complete set of partial 
differential fluid equations (with the variation of flow parameters in both radial and vertical direction) without 
assuming the vertical hydrostatic equilibrium, with the explicit inclusion 
of the vertical velocity (representing outflow) and all the  
relevant components of the stress 
tensor apart from the usual $W_{r\phi}$, from the first principle. They used their solutions 
to study two extreme cases of the geometrically thick 
advective accretion flows: super-critical and high sub-critical, which are more probable 
regimes of the strong outflows and jets. They found that 
the flow parameters of the accretion-induced 
outflow/jet strongly depend on Shakura \& Sunyaev viscosity parameter $\alpha$ \cite{SS73} and cooling factor $f$. 

In the present paper, we use and extend the above mentioned work \cite{GM09} 
to study the detailed energetics of the accretion-induced 
outflow and then plausible jet. In order to do that, we stick to two extreme regimes 
of the black hole accretion: super-critical and high sub-critical, to describe ultra/highly luminous and under-luminous sources 
respectively. For obvious reasons, we do not
repeat the calculation and the formulation of the disc-outflow model given
earlier \cite{GM09}, but will recall them appropriately.

We arrange the paper in the following manner. In the next 
section, we formulate the equations for the energetics of the 
accretion-induced outflow. In \S 3, we study the properties of the disc-outflow energetics. 
Finally we end in \S 4 with a discussion and implications. 

\section{Energetics of the accretion-induced outflow}

In order to study the energetics of the outflow, we need to compute the 
mass outflow rate. Blandford \& Begelman (1999) generalized the ADAF solution \cite{NY94}
by including wind and outflow, 
assuming the mass inflow rate to be proportional to 
$r^{p}$, where $0 \leq p <1$ ($r$ be the radial coordinate). Chakrabarti and his collaborators \cite{DAS, CHAK} 
also attempted to calculate the mass outflow rate 
from a disc involving a shock without 
including the vertical flow explicitly. 
In the present paper, we derive this self-consistently. 
Integrating the continuity equation \cite{GM09} 
vertically about the equatorial plane we obtain 
\begin{eqnarray}
\frac{d}{dr} \int_{h_0}^{h} 4 \pi r \rho  v_r \, dz
\, + \, 4 \pi r (\rho (h) v_z (h)- \rho (h_0) v_z (h_0))  = 0, 
\label{1a}
\end{eqnarray}
where $v_r$ and $v_z$ are the radial and the vertical velocity respectively, $h(r)$ is an arbitrary disc scale height and 
$h_0$ the minimum (finite) height of the disc-outflow system, which is 
different for super-Eddington and sub-Eddington accretion flows, $z$ be the vertical coordinate. 
Here we deliberately focus on the disc-coronal region (sub-Keplerian halo),
from where the mass loss takes place in the form of wind or outflow, and hence the solution at 
equatorial plane looses significance.
The integral in the first term of the left-hand side of eqn. (\ref{1a}) 
represents the disc mass accretion rate given by
\begin{eqnarray}
\dot M_a (r) =  -\int_{h_0}^{h} 4 \pi r \rho  v_r \, dz.
\label{1b}
\end{eqnarray}
The second term then can be attributed to the rate of change of outflow (and then plausible jet) 
mass, given by  
\begin{eqnarray}
\frac{d \dot M_j (r)}{dr} = - 4 \pi r (\rho (h) v_z (h) - 
\rho (h_0) v_z (h_0)) = - \frac{d \dot M_a (r)}{dr}.
\label{1c}
\end{eqnarray}
Here, $\dot M_j (r)$ is analogous to the mass outflow rate 
represented as
\begin{eqnarray}
\dot M_j (r) \, = \, - \int 4 \pi r (\rho (h) v_z (h) - \rho (h_0) v_z (h_0)) \, dr + c_j,
\label{1d}
\end{eqnarray}
where the constant $c_{j}$ is to be determined by an appropriate boundary condition.
Thus from eqn. (\ref{1c}) the total mass accretion rate can be written as 
\begin{eqnarray}
\dot M=\dot M_a (r) \,+ \, \dot M_j (r).
\label{1f}
\end{eqnarray}
Eqn. (\ref{1f}) entails that under the stationary condition the radial 
mass flux in the disc decreases as the inflowing matter approaches 
the central object, at the same rate at which the vertical mass flux increases to 
maintain a constant $\dot M$ which is exactly the net mass accretion rate 
at infinity. 

To compute the power of the outflow and then plausible jet, 
following previous work \cite{GM09}, we integrate the disc-energy conservation equation 
for an accretion-induced outflow over the disc scale height which yields 
\begin{eqnarray}
\frac{d}{dr} \int_{h_0}^{h} 4 \pi
 r {\mathcal F_r} \, dz
\, + \, 4 \pi r ({\mathcal F_z (h)} - {\mathcal F_z (h_0)}) \, = \, 0. 
\label{2a}
\end{eqnarray}
Here, ${\mathcal F_z}$ is the vertical component of the total energy flux given by 
\begin{eqnarray}
{\mathcal F_z} = \biggl(\frac{v^2}{2} + \frac{\gamma}{\gamma-1} 
\frac{P}{\rho} + \phi_G  \biggr) \rho v_z - v_{\phi} W_{\phi z} + F_z, 
\label{2b}
\end{eqnarray} 
where $v^2=v_r^2+ v^2_{\phi}+ v_z^2$, $\phi_G$ is the gravitational
potential, $F_z$ the radiative flux from 
the disc surface, $W_{\phi z}$ the $\phi z^{\rm th}$ component of the stress tensor, $\gamma$ 
the ratio of the specific heats of the gas-radiation mixture. 
We thus obtain the power of the outflow from eqn. (\ref{2a}), 
which is the total power removed from the disc by the outflow, as  
\begin{eqnarray}
P_j (r) = \int 4 \pi r ({\mathcal F_z (h)} - {\mathcal F_z (h_0)}) \, dr.
\label{2c}
\end{eqnarray}
We then calculate the disc luminosity in presence of outflow (and jet) 
using disc-energy conservation equation described 
earlier \cite{GM09} 
\begin{eqnarray}
L=(1-f)\int \left(\int^{h}_{h_{0}} Q^{+} 4 \pi r dz\right) dr, 
\label{2d}
\end{eqnarray}
where $Q^{+}$ is the total viscous heat generated in the disc. 
We now introduce three dimensionless parameters which correlate the disc and 
the outflow leading to jet: $q_{jm} (r) = {\dot M_j (r)}/{\dot M}$, 
$q_{jp} (r)= {P_j (r)}/{\dot M c^2}$ and $q_l={L}/{\dot M c^2}$. We do not 
describe here the generalized 
Bernoulli's number $(B_E)$, explicitly shown earlier \cite{GM09}, 
which represents the total energy of the system. 

\section{Properties of the energetics of disc and outflow}

Accretion by a black hole (or any other central star) 
is the primary source of the mass outflow and jet
formed in the inner hot region of the 
disc. At the first instant we neglect the contribution of magnetic field which is likely to be the origin of 
collimation and acceleration of jet. However, doubts can be raised about the significance of the magnetic field  for production, 
collimation and acceleration of jets in ULXs and highly luminous AGNs and quasars, which are super-critical and highly 
radiation trapped systems (see Fabrika 2004, for details).  
We argue that the radiation pressure is 
likely to be the plausible reason for strong outflows and then plausible jets in high 
mass accretion flows \cite{LOV94}. The inflowing matter is expected to be ejected and accelerated through a 
funnel like region \cite{FUK87, FAB} by the strong radiation pressure leading to strong outflow and then plausible jet from 
the disc. In early, based upon the geometrically 
thick accretion disc model, it was shown that very 
narrow and deep funnels are formed around the rotation axis of the accretion disc and most of the 
disc energy flux is radiated from the surface of the funnels. The radiation pressure may 
accelerate the outflowing mass in funnels up to relativistic velocities in the form of two jets pointing in 
opposite directions as observed in SS433 \cite{JAP80, AP80}. 
In the inner accretion disc, a few Schwarzschild radii away from the black hole, 
as the infall time scale is much smaller than the viscous time scale to
transport angular momentum outside, matter has no time to loose angular
momentum and attains a near constant value. Around
this region the centrifugal force becomes comparable to the gravitational force, 
and thus the incoming matter slows down at this centrifugally dominated
region and gets puffed up \cite{CHAK}. Therefore, a funnel
type geometry is likely to form around the 
region and the radiation pressure blows up the matter through
them.

However, the magnetic field 
is likely to play the role in generating turbulence in the hot disc systems \cite{BH91}. In fact, for critical 
or sub-critical accretion flows, the jet is likely to form due to the magnetic 
activity in the disc \cite{BP82, CAM}. We do not intend to discuss here the possible reason of the emanation 
of the jet. Whatever the explanation/mechanism be, the potential 
dynamics of the accretion-induced outflow and then plausible jet always upholds the 
conservation laws which shows several 
interesting features. The dimensionless 
parameters $q_{jm} (r)$, $q_{jp} (r)$, $q_l$ and $B_E$ 
carry the information 
of the disc-outflow energetics. In the next two subsections 
we analyse their features for super-critical and 
sub-critical accretion flows respectively. 

\subsection {Super-critical accretion}

To study the high mass accretion flows, we 
consider $\dot M \sim 3 \times 10^{-4} {M_{\odot}}/{yr}$ 
corresponding to the central mass $M \sim 10 M_{\odot}$, which is 
a probable estimate of the accretion rate 
of our own galactic super-critical accretor SS433 \cite{FAB}. 
We express $\dot M$ in terms of critical Eddington 
unit so that, for our case, $\dot M \sim 10^4 \dot M_{cr}$. This 
makes the energetic profiles independent of mass of 
the central star and one can extend our model to the case of high or 
ultra-luminous AGNs and quasars. Figures 1a and 1b depict
the variation of mass outflow rate and power of the 
outflow and then plausible jet respectively as functions of $r$. 
The nature of profiles is 
similar to that of vertical component of velocity 
$v_z$, explained in detail earlier \cite{GM09}. 
At $f=0.4$, a substantial amount of 
matter and radiation is extracted 
in the form of strong outflow leading to jet and thus the 
power of the outflow is 
significantly high, which falls off rapidly for $f > 0.4$. 

The luminosity profile (Fig. 1c) exhibits an 
interesting feature. At small $r$, it increases slowly (although not
very clear from the figure that is given in the logarithmic scale) 
with $r$ and eventually attains a near constant value. At 
high $\alpha$, the luminosity corresponds to the Eddington or the super-Eddington
accretion. However the system becomes sub-Eddington at 
low $\alpha \sim 0.01$, even if $\dot M \sim 10^4 \dot M_{cr}$ and $f$ low. This extreme sensitivity is due 
to the fact that $\alpha$ directly controls the 
viscous heat generation. At $f = 0.4$, for 
$\alpha \sim 0.3$, $L$ reaches $\sim 10^{41} erg/s$, resembling 
highly luminous and ultra-luminous XRBs 
(e.g. CXOU~J095550.2+694047 \cite{KAR}, GRS~1915+105, GRO~J1655-40 
\cite{MIRR98}). At $f > 0.4$, $L$ decreases in orders of magnitude 
but still lies within the range of the high or ultra-luminous regime 
(as in the case of SS433). 
For a black hole of mass $M \sim 10^{6} - 10^{9} M_{\odot}$, it can 
be shown that $L$ attains a value $\sim 10^{46} - 10^{49}$ erg/s 
for $\alpha = 0.3$ and $f = 0.4$. Such luminosity is commonly 
observed in highly luminous AGNs and ultra-luminous quasars 
(e.g. PKS 0743−67), possibly in ULIRs \cite{GEN} and narrow-line Seyfert 1 galaxies (e.g. \cite{MINS}). 
The nature of the variation of $B_E$ has been shown in Fig. 1d,e,
already explained by Ghosh \& Mukhopadhyay (2009) with three-dimensional plots. 
 
The truncation of the curves attributes to the truncation of the disc due to evaporation to 
corona. With the decrease 
of $f$, which corresponds to more increase of radiation pressure, the disc gets more radiation pressure dominated and the 
matter gets blown up due to the strong radiation pressure further out. 
At this situation, the disc tends to become more Keplerian and thus
centrifugally dominated. Therefore the centrifugal
dominated boundary layer is likely to form at a larger radius. 
As a result the outflow rate and then the outflow/jet power 
increase significantly and the disc gets 
truncated at a larger radius. 

In Fig. 2, we show a comparison of $v_z, q_{jm}$, $q_{jp}$ and $q_l$ 
between two values of $\gamma$ at the same set of $\alpha$ and $f$. With a small 
increase in the value of $\gamma$ from $1.4$ to $1.444$, the above parameters 
decrease by an enormous margin.  The abrupt change of the corresponding 
parameters reflects that the system is very sensitive to $\gamma$ and then
$\beta$ which is the ratio 
of the gas pressure to total pressure \footnote {The relation between $\gamma$ and $\beta$ is given in Ghosh \& Mukhopadhyay 2009.}. 
An increase of $\gamma$ from $1.4$ to $1.444$ corresponds to an increase 
of $\beta$ from $0.3$ to $0.5$, which further corresponds to the conversion from 
radiation dominated to marginally gas dominated system. This signifies that 
the gas pressure in the system increases, which means that the radiation 
pressure of the system suddenly 
decreases. The decrease in the radiation pressure inhibits strong outflow
\footnote{In super-critical, radiation pressure dominated regime, strong 
radiation pressure blows up the matter to form outflow and consequently plausible jet.}. Hence the power of the outflow and mass outflow rate fall in order of 
magnitudes. The same reasons are applicable in order to explain the variation 
of luminosity profiles. 

\subsection {Sub-critical accretion}

Here we consider that the flow has a mass accretion 
rate $\dot M \sim 10^{-2} \dot M_{cr}$ 
corresponding to $M \sim 10^6 - 10^7M_{\odot}$, which renders the flow to be thick and
gas pressure dominated. Figure 3 shows that the 
mass outflux and the power extracted from the disc by the outflow
are much less compared to that of the high mass accretion flows. Even for a near 
extreme $f \sim 0.7$, $B_E$ is $\sim 4$ times larger for super-critical flows 
compared to the ideal case of an advection 
dominated $(f \rightarrow 1)$ sub-critical accretion flow. This signifies that outflows are more
probable for super-Eddington accretion flows compared to that of sub-Eddington flows. 
For low $\alpha$ 
and $f=0.9$, we obtain a luminosity $L \sim 10^{33} erg/s$ corresponding to 
$M \sim 10^6 M_{\odot}$ (Fig. 3c). For high $\alpha \sim 0.3$, $L$ increases 
by three orders of magnitude (not shown in the figure) at the same $f$. 
Such low luminosities 
are observed in many under-luminous AGNs 
(e.g. Sagittarius $A^{*}$; \cite{MAH}). 

\section{Discussion}

Based on a 2.5 dimensional hydrodynamical formulation of a 
disc-outflow coupling system with a set of self-similar solutions
\cite{GM09}, we have studied the energetics of the system. 
The self-similar approach is not just a mere tool to solve the non-trivial 
coupled partial differential equations. The accretion flow indeed may 
exhibit self-similar behaviour. 
It was shown earlier that the black hole system GRS~1915+105 is chaotic in nature \cite{MISRA} 
that supports the idea of inner disc instability and then turbulence.
It was also found that the corresponding correlation/fractal dimension to be similar
to that in the Lorenz system which is a model example of an ideal chaos.
The low correlation/fractal dimension implies possible self-similarity into the system. 
In reality the accretion flows with 
strong outflows are likely to be geometrically thick and advective, which 
more possibly occur in ultra-luminous and under-luminous accreting 
sources. The standard Keplerian model \cite{SS73} fails to describe these two opposite traits of observational 
signatures, which are possible sources of powerful outflows and jets. 

The energetic profiles strongly depend on $\alpha$ and $f$. They also 
reveal that outflows and jets are more probable and powerful for 
super-Eddington accretion flows compared to the 
sub-critical ones, 
which correspond to ultra-luminous and under-luminous sources respectively. 
For super-Eddington flows (Fig.1), the disc gets truncated at larger 
radii compared to the sub-Eddington ones. This means that the outflow/jet 
may occur at the inner region of the disc and the disc-corona transition 
region may form in the vicinity to the central star in case of sub-Eddington flows. 
However, a slight departure of $\gamma$ from $\sim 1.4$ to 
a higher value leads to an abrupt decrease in the energetics of the 
flow. We have shown that at super-Eddington accretion flows $(\dot M \sim 10^4 \dot M_{cr})$, 
with an appropriate choice of $\alpha$ and $f$, the luminosity calculated 
from our model is $L \sim 10^{41}$ erg/s corresponding to a black hole of mass 
$M \sim 10 M_{\odot}$. Such a high luminosity is commonly 
observed in exotic ULXs 
in external galaxies or in our own super-critical accretor SS433. 

With the appropriate 
choice of free parameters $\alpha, f, M$ and $\dot M$, we 
have shown that our model can reproduce 
the luminosities observed in ultra luminous sources and other family of high luminous AGNs 
(Fig. 1c), 
which probably accrete super-critically. 
Similarly, the luminosity observed in under-luminous 
AGNs (e.g. Sagittarius $A^{*}$) can also be reproduced by our model at $f \rightarrow 1$
(Fig. 3c). This particular situation is similar to ADAF \cite{NY94}. The limitation of ADAF can be addressed 
by incorporating outflow from the disc (as of Ghosh \& Mukhopadhyay 2009), 
which may turn the disc into a non-radiative accretion flow 
and thus can explain luminosity profile of under-luminous sources. 
As our solutions can explain the observed luminosities of both ultra-luminous sources and 
under-luminous AGNs with the appropriate choice of free parameters, 
we should put our model to more observational tests. Although we have neglected the contribution 
of the magnetic field, whose importance for super-critical accretion flows is 
probably not significant \cite{GM09}, 
the inclusion of the magnetic field will render the system to be more realistic. 
Such a work will be pursued in the course of time. In the follow up work, we plan 
to analyse a full scale numerical solution
to conform/verify the results discussed here.

\section*{Acknowledgments}
This work is partly supported by a project, Grant No. SR/S2HEP12/2007, funded
by DST, India. \\



\begin{figure}
\centering
\includegraphics[width=1.0\columnwidth]{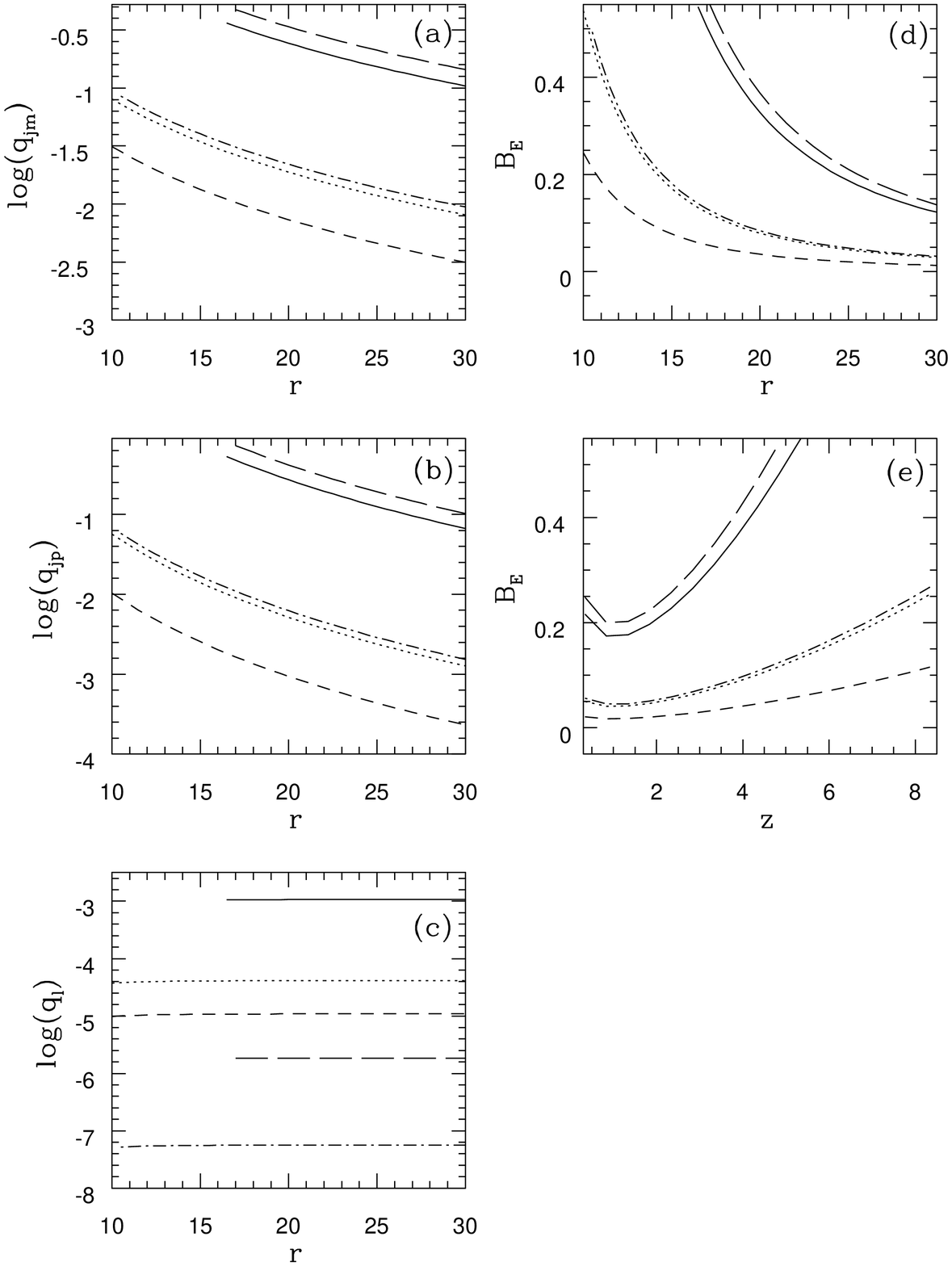}
\caption{
Variation of (a) mass outflow rate, (b) outflow/jet power, (c) luminosity in 
unit of $\dot M c^2$, (d) Bernoulli's constant, as functions of radial
coordinate for super-Eddington accretion flows. (e) Variation 
of Bernoulli's constant as a function of $z$. 
Solid, dotted, dashed curves are for $\alpha = 0.3$ and
$f = 0.4, 0.5, 0.7$ respectively. Long-dashed, dot-dashed 
curves represent flow with $\alpha = 0.01$ and $f = 0.4, 0.5$ respectively.
Other parameters are $\gamma = 1.4$, corresponding $\beta \sim 1/3$, and 
$z = 5$.
 }
\label{fig1}
\end{figure}


\begin{figure}
\centering
\includegraphics[width=0.80\columnwidth]{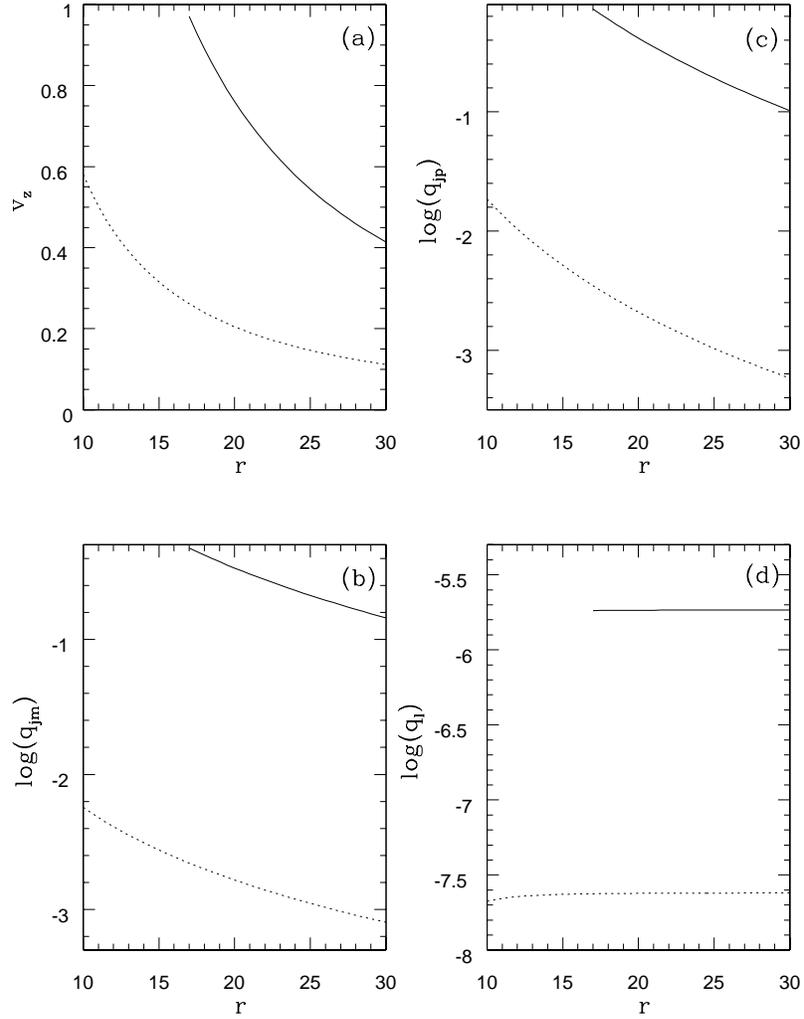}
\caption{
A comparison of various quantities between two different $\gamma$ 
for super-Eddington accretion flows as functions of radial coordinate. 
Solid and dotted curves are for $\gamma = 1.4$ and $\gamma = 1.44$ 
respectively. Other parameters are $\alpha = 0.01, f = 0.4, z = 5$. 
 }
\label{fig2}
\end{figure}


\begin{figure}
\centering
\includegraphics[width=1.0\columnwidth]{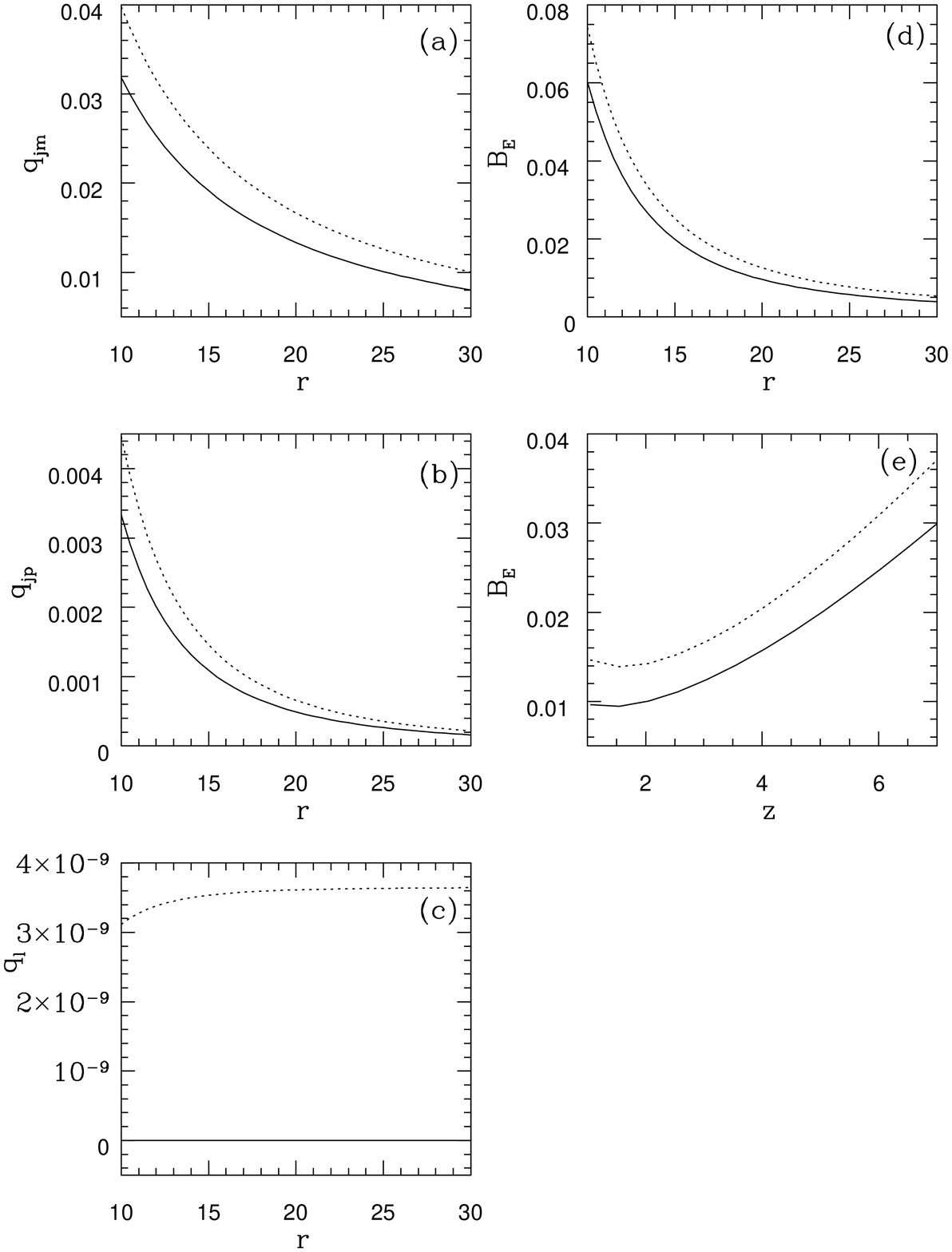}
\caption{
Same as Fig. 1, but for sub-Eddington accretion flows. 
Solid and dotted curves are for $\alpha = 0.3, f = 1$ and
$\alpha = 0.01, f = 0.9$ respectively. 
Other parameters are $\gamma = 1.6$, corresponding $\beta \sim 0.89, z = 5$.
 }
\label{fig3}
\end{figure}


\begin{thebibliography}{}
\bibitem[Abramowicz \& Piran 1980]{AP80} M. Abramowicz, \&  T. Piran, ApJ, 241, 7 (1980)
\bibitem[Balbus \& Hawley 1991]{BH91} S. A. Balbus, \&  J. F. Hawley, ApJ, 376, 223 (1991)
\bibitem[Begelman, Blandford \& Rees 1984]{BBR} M.C. Begelman, R. D Blandford, \&  M.J. Rees, RvMP, 56,
255 (1984)
\bibitem[Begelman et al. 2006]{BEG}  M. C. Begelman,  A. R. King, \&  J. E. Pringle, 
MNRAS, 370, 399 (2006)
\bibitem[Blandford \& Begelman 1999]{BB99} R. D. Blandford,  \&  M. C. Begelman, MNRAS, 303, 1 (1999)
\bibitem[Blandford \& Payne 1982]{BP82} R. D. Blandford, \&  D.G. Payne, MNRAS, 199, 883 (1982)
\bibitem[Blandford \& Rees 1974]{BR} R. D. Blandford, \&  M. J Rees, MNRAS, 169, 395 (1974)
\bibitem[Blandford \& Znajek 1977]{BZ} R. D. Blandford, \&  L. Znajek, MNRAS, 179, 433 (1977)
\bibitem[Camenzind 1986]{CAM} M. Camenzind, A\&A, 156, 137 (1986) 
\bibitem[Chakrabarti 1999]{CHAK} S. K. Chakrabarti, A\&A, 351, 185 (1999) 
\bibitem[Das \& Chakrabarti 1999]{DAS} T. K. Das, \& S. K. Chakrabarti, CQG, 16, 3879 (1999) 
\bibitem[Fabbiano 2004]{FAB} G. Fabbiano, RMxAC, 20, 46 (2004)
\bibitem[Fabrika 2004]{FAB} S. Fabrika, ASPRv, 12, 1 (2004) 
\bibitem[Ferrari 1998]{FER} A. Ferrari, ARA\&A, 36, 539 (1998)
\bibitem[Fukue 1987]{FUK87} J. Fukue, PASJ, 39, 679 (1987)
\bibitem[Genzel et al. 1998]{GEN} R. Genzel et al., ApJ, 498, 579 (1998)
\bibitem[Ghosh \& Mukhopadhyay 2009]{GM09} S. Ghosh, \&  B. Mukhopadhyay, RAA, 9, 157 (2009)  
\bibitem[Hawley \& Balbus 2002]{BL20} J. F. Hawley, \&  S. A. Balbus, ApJ, 573, 738 (2002) 
\bibitem[Hawley \& Krolik 2006]{HK06} J. F. Hawley, \&  J. H. Krolik, ApJ, 641, 103 (2006)
\bibitem[Jaroszy\'nski, Abramowicz \& Paczy\'nski 1980]{JAP80} M. Jaroszy\'nski, M. A. Abramowicz, \&  B. Paczy\'nski, AcA, 30, 1 (1980)
\bibitem[Kaaret et al. 2001]{KAR} P. Kaaret et al., MNRAS, 321, 29 (2001)
\bibitem[Lipunova 1999]{LIP} G. V. Lipunova, AstL, 25, 508 (1999)
\bibitem[Lovelace et al. 1994]{LOV94} R. V. E. Lovelace,  M. M. Romanova, \& 
W. I. Newman, ApJ, 437, 136 (1994)
\bibitem[Mahadevan 1998]{MAH} R. Mahadevan, Nature, 394, 651 (1998) 
\bibitem[Margon 1984]{MAR84} B. Margon, ARA\&A, 22, 507 (1984)
\bibitem[Migliari \& Fender 2006]{MF06} S. Migliari, \&  R. P. Fender, MNRAS, 366, 79 (2006)
\bibitem[Mineshige et al. 2000]{MINS} S. Mineshige,  T. Kawaguchi,  M. Takeuchi,  \&  K. Hayashida, PASJ, 52, 499 (2000) 
\bibitem[Mirabel 2003]{MIR} I. F. Mirabel, New Ast. Rev., 47, 471 (2003)
\bibitem[Mirabel \& Rodriguez 1994]{MIRR94} I. F. Mirabel, \&  L. F Rodriguez,  Nature, 371, 46 (1994)
\bibitem[Mirabel \& Rodriguez 1998]{MIRR98} I. F. Mirabel, \&  L. F Rodriguez,  Nature, 392, 673 (1998) 
\bibitem[Misra et al. 2004]{MISRA} R. Misra,  K. P. Harikrishnan,  B. Mukhopadhyay,  G. Ambika,  \&  A. K. Kembhavi, ApJ, 609, 313 (2004)
\bibitem[Mizuno et al. 2006]{MIZ} Y. Mizuno,  K-I. Nishikawa,  S. Koide,  P. Hardee, \&  G. J. Fishman, arXiv:astro.ph.9344 (2006)
\bibitem[Mundt 1985]{MUN85} R. Mundt, Obs, 105, 224 (1985)
\bibitem[Mushotzky 2004]{MUS} R. Mushotzky, Prog. Theor. Phys. Suppl., 155, 27 (2004) 
\bibitem[Narayan \& Yi 1994]{NY94} R. Narayan,  \&  I. Yi, ApJ, 428, 13 (1994) 
\bibitem[Ohsuga et al. 2005]{OSH} K. Ohsuga, M. Mori, T. Nakamoto, S. Mineshige, ApJ, 628, 368 (2005)
\bibitem[Pudritz \& Norman 1986]{PN86} R. E. Pudritz, \&  C. A. Norman, ApJ, 301, 571 (1986)
\bibitem[Punsly \& Tingay 2005]{PT} B. Punsly,  \&  S. J. Tingay, ApJ, 633, 89 (2005) 
\bibitem[Shakura \& Sunyaev 1973]{SS73} N. Shakura,  \&  R. Sunyaev, A\&A, 24, 337 (1973)
\bibitem[Ustyugova et al. 1999]{UST} G. V. Ustyugova,  A. V. Koldoba,  M. M. Romanova,  V. M. Chechetkin,  \& R. V. E.  Lovelace, ApJ, 516, 221 (1999)


\end{thebibliography}
\end{document}